\RequirePackage{fix-cm}
\documentclass[twocolumn]{svjour3}          
\smartqed  

\usepackage{graphicx}
  \usepackage{dcolumn}
  \usepackage{bm}
  \usepackage{ulem}
  \usepackage[utf8]{inputenc}
  \usepackage[english]{babel}
  \graphicspath{{figs/}{figs:}{\figs}}
  \usepackage[usenames,dvipsnames]{color}
  \usepackage[]{ulem}
  \usepackage[colorlinks=true,
  linkcolor=blue,
  urlcolor=blue,
  citecolor=blue,
  breaklinks=true]{hyperref}
  \usepackage{xcolor}
  \newcommand{\comment}[1]{\textcolor{red}{#1}}
  \renewcommand{\comment}[1]{\relax}
  \newcommand{\todelete}[1]{\textcolor{green}{\sout{#1}}}
  \renewcommand{\todelete}[1]{\relax}

\begin{document}

\title{Instability of a horizontal water half-cylinder under vertical vibration
}


\author{Dilip Kumar Maity$^1$  \and            
        Krishna Kumar$^{1,\dagger}$      \and 
        Sugata Pratik Khastgir$^1$ 
}  
\institute {$^1$ Department of Physics, Indian Institute of Technology Kharagpur, Kharagpur-721302, India \\
$\dagger$ E-mail: {kumar.phy.iitkgp@gmail.com} 
}

\date{\today}

\maketitle

\begin{abstract}
We present the results of an experimental investigation on parametrically driven waves in a water half-cylinder on a rigid horizontal plate, which is sinusoidally vibrated in the vertical direction. As the forcing amplitude is raised above a critical value, stationary waves are excited in the water half-cylinder. Parametrically excited subharmonic waves are non-axisymmetric and qualitatively different from the axisymmetric Savart-Plateau-Rayleigh waves in a vertical liquid cylinder or jet. Depending on the driving frequency, stationary waves of different azimuthal wave numbers are excited. A linear theory is also supplemented, which captures the observed dispersion relations quantitatively.
\keywords{Interfacial instability \and Parametric excitation \and Vibrating water half-cylinder}
\end{abstract}

\section{Introduction}
The parametric excitation of stationary waves on the free surface of an extended layer of liquid is known since the pioneering work of Faraday~\cite{Faraday_1831}. The excited standing waves are subharmonic, i.e., the frequency of such waves is equal to half of the driving frequency. These waves may also be synchronous with the driving in a thin viscous layer~\cite{Kumar_1996}. Synchronous parametric waves were also observed in experiments~\cite{Mueller_etal_1997}. Faraday instability leads to interesting fluid patterns~\cite{Faraday_patterns}. On the other hand, capillary instability of a vertical liquid cylinder or a jet is known since the pioneering experiments of Savart~\cite{Savart_1833} and Plateau~\cite{Plateau}. Rayleigh investigated theoretically the instability of liquid jets~\cite{Rayleigh}. A vertical liquid cylinder or jet develops an axisymmetric bead like structure, which ultimately leads to breaking of the jet into detached liquid drops~\cite{Lamb_1932}. Plateau remarked that a liquid jet was stable for all  purely non-axisymmetric deformations, but was unstable for axisymmetric varicose deformations with wavelengths exceeding the circumference  of the cylinder~\cite{Chandrasekhar_1961}. Other experiments~\cite{Donnely_Glaberson_1966,Moseler_Landman_2000} also confirm the instability of a vertical jet through axisymmetric perturbations. 
 
We present, in this article, results of an experiment that allows excitation of only non-axisymmetric waves on a horizontal water half-cylinder under sinusoidal vibration in the vertical direction. This novel experiment combines two fluid instability problems.  Faraday instability breaks the invariance under continuous time translation at instability onset and leads to excitation of subharmonic stationary waves. These waves are invariant under time translation by a period equal to double the period of external driving. On the other hand, Savart–Plateau–Rayleigh instability~\cite{Savart_1833,Plateau,Rayleigh} is strongly influenced by curvature of a liquid jet or a column. A long water half-cylinder has translational symmetry along one direction and its base sticks to plate due to no-slip condition. It is, therefore, qualitatively different from the classical Faraday experiment, the Savart-Plateau-Rayleigh problem and a vibrating spherical liquid drop~\cite{spherical_drop} problem. As the amplitude of driving is raised above a critical value in our experiment, depending on the driving frequency, three different types of subharmonic stationary waves are excited: half-bead like structure, waving half-cylinder and complex knitting patterns. These standing waves are non-axisymmetric, unlike the excited modes of a vertical liquid cylinder~\cite{Lamb_1932,Chandrasekhar_1961}. The dispersion curves show windows of frequencies where stationary waves are not sustained at the primary instability. An effective linear theory for this problem, explaining the gross features of the observed dispersion curves, is also presented .

\section{Experimental setup}

\begin{figure}[ht]
\begin{center}
\includegraphics[height=!, width=8.5 cm]{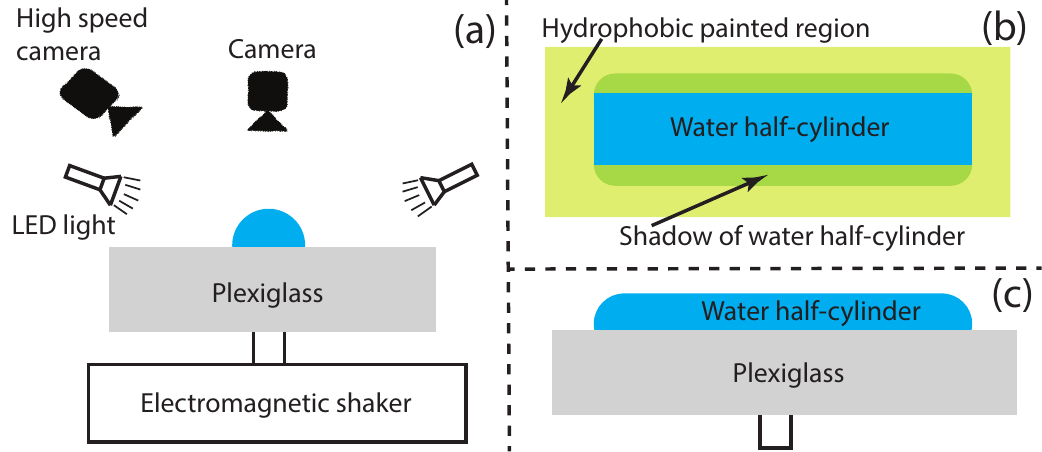}
\caption{(Color online): Schematic diagram of the experimental setup: 
	(a) Side view seen along the axis of the water half-cylinder. 
   	(b) Top view of the half-cylinder [blue (dark gray) color] surrounded by a hydrophobic painted region [yellow-green (gray) color] on the plate.
    (c) Side view from the direction normal to the cylinder axis.}
\label{schematic}
\end{center}
\end{figure}

\begin{figure}[]
\begin{center}
\includegraphics[height=!, width=8.5 cm]{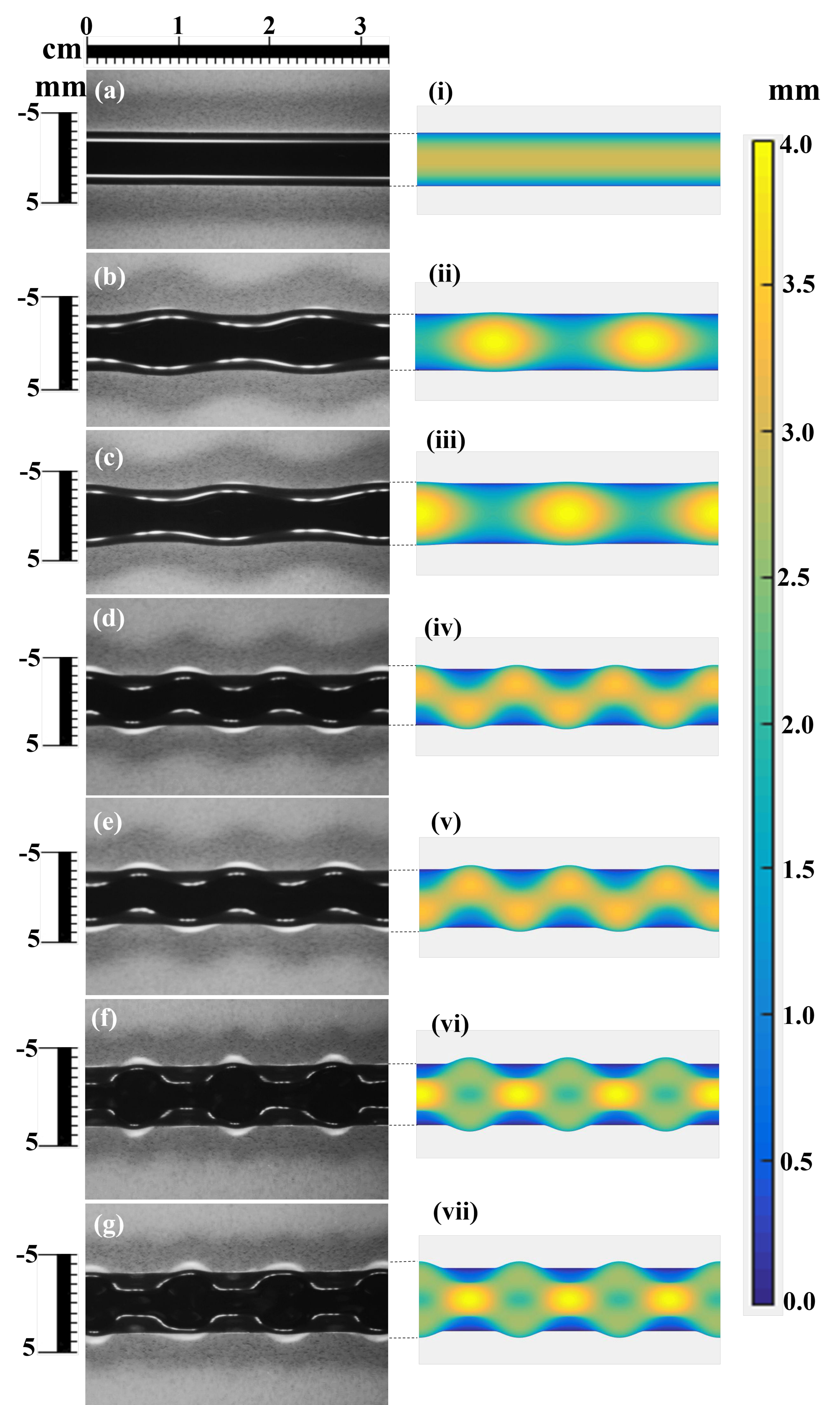}
\caption{(Color online): Top view of experimentally observed stationary waves on a portion of water half-cylinder ($L=10.0\pm 0.1$ cm, $R=3.1\pm 0.1$ mm) (see the left column) and computer generated patterns (right column): Stationary liquid cylinder [(a) \& (i)]; a chain of periodic half-beads (azimuthal wave number $m = 1$) on the liquid cylinder at some instant [(b) \& (ii)] and the same half the period later [(c) \& (iii)] for $f = 20$ Hz and $a = 0.56\pm 0.01 g$; waving half-cylinder ($m = 2$) at an instant [(d) \& (iv)] and the same half the period later [(e) \& (v)] for $f = 60$ Hz and $a = 0.85\pm 0.01 g$; complex knitting pattern ($m = 3$) at an instant [(f) \& (vi)] and the same half the period later [(g) \& (vii)] for $f = 90$ Hz and $a = 2.48\pm 0.01 g$. The white broken curved lines on the patterns captured in experiments are reflections of light from the curved surface of water half-cylinder. Scales along the axis of the half cylinder and along its diameter are shown in the left column. The color scale in the right column shows the height of the water half-cylinder. The yellow (white) color stands for the maximum height and dark blue (black) color stands for the minimum height in a pattern. The straight dotted lines between two columns compare the width of computer generated patterns  with  the corresponding images of actual water half-cylinder captured in experiments.
} \label{top_view}
\end{center}
\end{figure}

A schematic diagram of the experimental setup is shown in Fig.~\ref{schematic}. The flat surface of a square plexiglass plate ($20 \times 20$ cm$^2$) of thickness $1$ cm is coated with a super-hydrophobic paint (Rust-Oleum 275619 NeverWet Nano) except for a rectangular region in the middle. The thickness of coating of the hydrophobic paint was measured using a screw gauge. It is $40\pm 10$ microns. The opposite surface of the plexiglass plate was painted black to remove reflections from the bottom of the plate. This plate was then rigidly fixed to the vibrating base of an electromagnetic shaker [Model no. V350, Data Physics Corporation (DPC)].  A vibration controller (SignalStar Scalar Vibration Controller 2.4.998, DPC) connected to a power amplifier (DSA5-2K, DPC) sets the amplitude and frequency of the electromagnetic shaker. A small accelerometer (256HX-10, Isotron) was attached to the base of the plate and connected to the vibration  controller unit to monitor the acceleration amplitude of the vibrating plate. A sufficient amount of distilled water was poured on the unpainted region of the plate drop by drop using a syringe so that a horizontal water half-cylinder was formed. The hydrophobic paint outside water half-cylinder pinned its bottom boundary on the plexiglass plate. The height of enclosed water in the unpainted rectangular region on the plate varied from $2.0$ to $4.5$ mm, which was $50$ to $110$ times larger than the thickness of the hydrophobic coating on the plate. To maintain the shape and size of the water half-cylinder against evaporation during the experiment, its shadow length was monitored using a camera (Basler scA1000-30gm) fixed above the water half-cylinder. A few drops of distilled water were added every $10$ minutes to compensate the water evaporation. The temperature of the laboratory during the experiment was maintained at $24\,^{\circ}{\rm C}\pm 1\,^{\circ}{\rm C}$. Diffused LED lighting was used for illumination from two sides of the half-cylinder. The dynamics of fluid patterns was captured using a high speed camera (Chronos 1.4, Kron Technologies, Canada), which has a resolution of $1280 \times 360$ at the rate of 2999 fps. The camera was so placed that its axis made an angle $\theta = 60^{\circ}$ or $40^{\circ}$ with the horizontal plane. 

The wavelengths of different stationary waves were extracted from captured images using a software~\cite{de-Jesus_2017,Tracker} known as Tracker 4.11.0,  which is a free video analysis and modeling tool built on the Open Source Physics (OSP) Java framework.  First the measuring tools of Tracker software were calibrated and validated with the image of a measuring scale kept beside the stationary water half-cylinder. The calibrated measuring tools were then used to determine the wavelengths from different images of waves on water half-cylinder.

\begin{figure*}[]
	\begin{center}
		\includegraphics[height=!, width=17 cm]{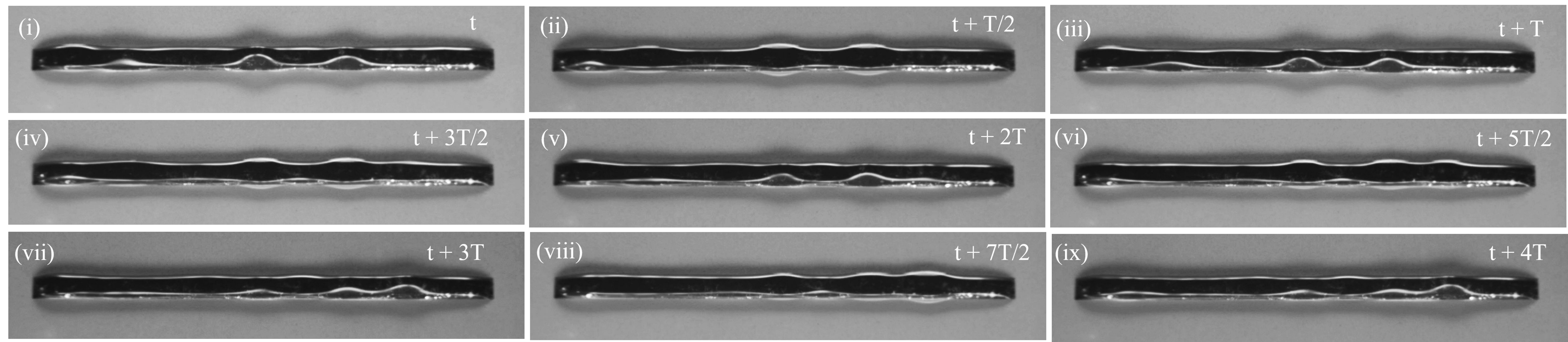}
		\caption{Experimentally observed irregular spatio-temporal patterns on the water half-cylinder of length $L=10.0\pm 0.1$ cm and radius $R=3.1\pm 0.1$ mm for $f= 34$ Hz and $a = 0.3\pm0.01 g$. Images [(i)-(ix)] are snapshots of the fluid patterns taken at an angle of $60^{\circ}$ from horizontal direction at a regular time interval $\tau = 1/(2f)$.  The white curved lines are reflections of light from the curved surface of water half-cylinder.}
		\label{sp_f34}
	\end{center}
\end{figure*}
\begin{figure*}[]
	\begin{center}
		\includegraphics[height=!, width=17 cm]{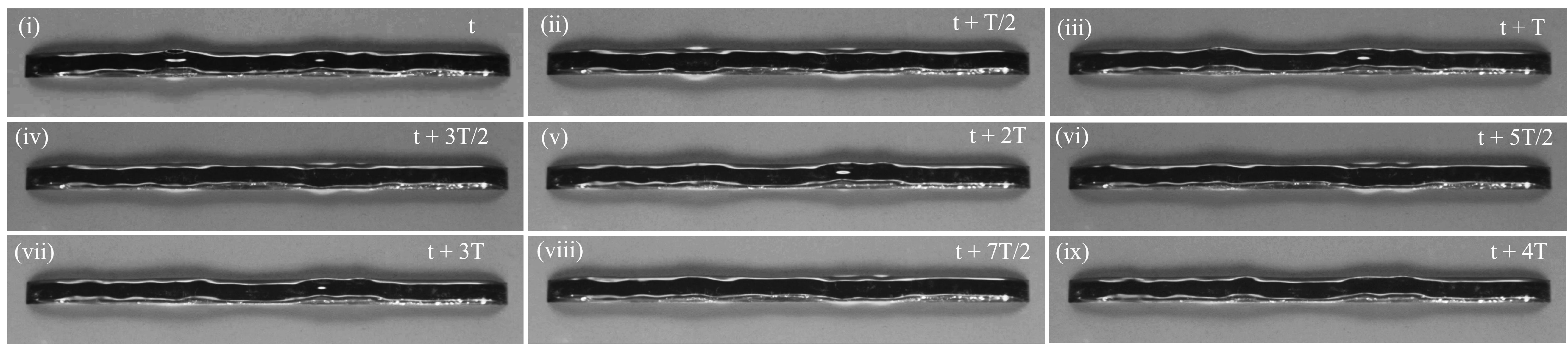}
		\caption{Irregular spatio-temporal patterns on the water half-cylinder of length $L=10.0\pm 0.1$ cm and radius $R=3.1\pm 0.1$ mm for $f= 80$ Hz and $a = 0.85\pm0.01 g$. Images [(i)-(ix)] are the snapshots of fluid patterns, taken at an angle of $60^{\circ}$ from horizontal direction, at a regular time ineterval of $\tau = 1/(2f)$. The white curved lines are due to  reflections of light from the curved surface of water half-cylinder.
}
\label{sp_f80}
\end{center}
\end{figure*}

\section{Results and discussion}
Experiments were performed for five different water half-cylinders of length $L$ ($10.0$ cm $\le L \le 15.0$ cm) and radius $R$ ($2.0$ mm $\le R \le 4.0$ mm) and repeated at least three times for each case. Qualitatively similar results were obtained in all the cases. The left column of Fig.~\ref{top_view} [(a)-(g)] displays the top view of experimentally captured images for a portion of the water half-cylinder of $L=10.0\pm 0.1$ cm and $R=3.1\pm 0.1$ mm. The scale is same for all the images. Figure~\ref{top_view}(a) shows the static water half-cylinder before the plate was subjected to vibration. As the bottom of the plate is painted black, the stationary water half-cylinder appears as black stripe. Two parallel white lines on the liquid half-cylinder are reflections from water surface of diffused LED lights used for illumination. The light gray portions on either side are the shadows of the water half-cylinder created by the LED light.

The acceleration amplitude $a$ of the oscillating plate was slowly raised in small steps at a fixed driving frequency $f$, and the excitation of waves was recorded by the camera. The driving frequency was raised in steps of $2$ Hz and the same procedure was repeated. The error in $f$ is $0.01$ Hz, which is negligibly small. As $a$ was raised above a critical value $a_\mathrm{c} (f)$ for a fixed value of $f$, the static water half-cylinder became unstable and stationary waves were excited. The stroboscopic light determined the frequency of excited stationary waves. The acceleration of the vibrating plate was controlled digitally using the vibration controller. The amplitude was raised in small steps of $0.01g$ and enough time was given for transients to die down. For very small values of $a$, there were waves of negligible amplitude without having clear structures. These waves had the frequency same as that of the driving. They were ignored, as they were not parametrically excited waves. To identify the critical acceleration $a_\mathrm{c} (f)$ of a parametrically excited subharmonic waves for a given value of the driving frequency $f$, the stroboscope was adjusted at a frequency $f/2$. As the driving amplitude was slowly raised, the stationary patterns started appearing above a certain value of the driving amplitude. This value is called critical amplitude $a_\mathrm{c} (f)$. This process was repeated three times at a particular frequency to estimate the average $a_\mathrm{c}$. This was also done by decreasing the driving amplitude in small steps from a value much above the $a_c$. No hysteresis was observed at the onset. The electromagnetic shaker can produce good sinusoidal vibrations for driving frequency above $15$ Hz. Parametrically excited subharmonic patterns were observed experimentally in the driving frequency range of $20$ Hz to $110$ Hz for all water half-cylinders considered here.

For a range of driving frequencies ($20$ Hz $\le f \le$ $110$ Hz), subharmonic (period, $\tau=2/f$) excitation of different stationary waves was observed. The water half-cylinder became a periodic chain of half-beads (horizontally chopped off) for a range of driving frequency from $20$ Hz to $28$ Hz. Figure~\ref{top_view}(b)-(c) shows the two phases of a portion of the chain of half-beads at two instants of time separated by half the period  of excited waves for $f= 20$ Hz and $a = 0.56\pm0.01 g$. The images clearly display the mirror symmetry about the vertical plane passing through the static cylinder axis. As the base of half-beads is fixed to the plate, the vertical oscillations make the transverse cross-sections of half-beads non-axisymmetric, although the mirror symmetry about the vertical plane through the symmetry axis is maintained. It is qualitatively different from axisymmetric beads observed  in a vertical liquid cylinder~\cite{Chandrasekhar_1961}. The axisymmetric mode is never excited in our case. 

Stationary waves were not observed for $30$ Hz $\le f \le 38$ Hz at the primary instability. Standing waves for a given value of azimuthal wave number $m$ are not excited, when the waves do not fit the water half-cylinder of finite length. It may not be easy to simultaneously acquire an appropriate azimuthal number and wavelength along its axis. The water half-cylinder showed a state of frustration and displayed irregular spatio-temporal dynamics at the primary instability in this case.  Figure~\ref{sp_f34} displays nine photographs of the water half-cylinder captured at a regular interval of the half the period of driving ($= T/2$).  A close look on these photographs shows that any two of them, separated by an interval of time $T$, are quite similar. They are, however, not exactly the same, which is quite clear by seeing the photographs at an interval of $3T$ or $4T$.  It appears similar to the phenomenon of superposition of waves with close multiple frequencies. These wave patterns seem to be quasiperiodic in time but aperiodic in space.

As $f$ was raised further, we observed again excitation of stationary waves for $40$ Hz $\le f \le$ $72$ Hz. The water half-cylinder started waving sub harmonically. The resulting pattern broke the mirror symmetry about the vertical plane through the cylinder axis. The waving cylinder showed the glide symmetry, i.e., the fluid pattern was invariant under a translation by $\lambda/2 = \pi/k$ along the cylinder axis followed by a mirror reflection about the vertical plane through the axis. Figure~\ref{top_view}(d)-(e) shows the two phases of the waving water half-cylinder at an instant and half the period later for $f= 60$ Hz and $a = 0.85\pm 0.01 g$.  

Stationary waves were not observed again in a frequency window between $74$ Hz and $84$ Hz, where the half-cylinder was in a state of frustration at the primary instability (see, Fig.~\ref{sp_f80}), as mentioned earlier. However, the quasiperiodic behavior is less pronounced than the earlier case. The waves are more irregular in space and time.  As the driving frequency was raised further, we observed subharmonically generated stationary waves in the form of a complex knitting pattern for $86$ Hz $\le f \le$ $110$ Hz.  Fig.~\ref{top_view}(f)-(g) show fluid patterns at two instants separated by half the wave period for $f=90$ Hz and $a = 2.48\pm 0.01 g$. The complex fluid pattern showed mirror symmetry about the vertical plane through  the cylinder axis. The complex patterns consist of mountain and valley like structures alternatively. The locations of mountains and valleys interchange periodically with time. 

Figure~\ref{sp_f34}(i)-(ix) shows the experimentally captured images of the water half-cylinder, at an angle of $60^{\circ}$ from the horizontal,  at different time instants for $f= 34$ Hz and $a = 0.3\pm0.01 g$. The fluid patterns  are no more periodic and the water half-cylinder is in a state of frustration. This happens whenever the stationary waves do not fit on a chosen liquid cylinder. Figure~\ref{sp_f80}(i)-(ix) also shows the images of of irregularly varying structures on the water half-cylinder  at different time instants for $f= 80$ Hz and $a = 0.85\pm0.01 g$.

\begin{figure}[t]
\begin{center}
\includegraphics[height=!, width=8.5 cm]{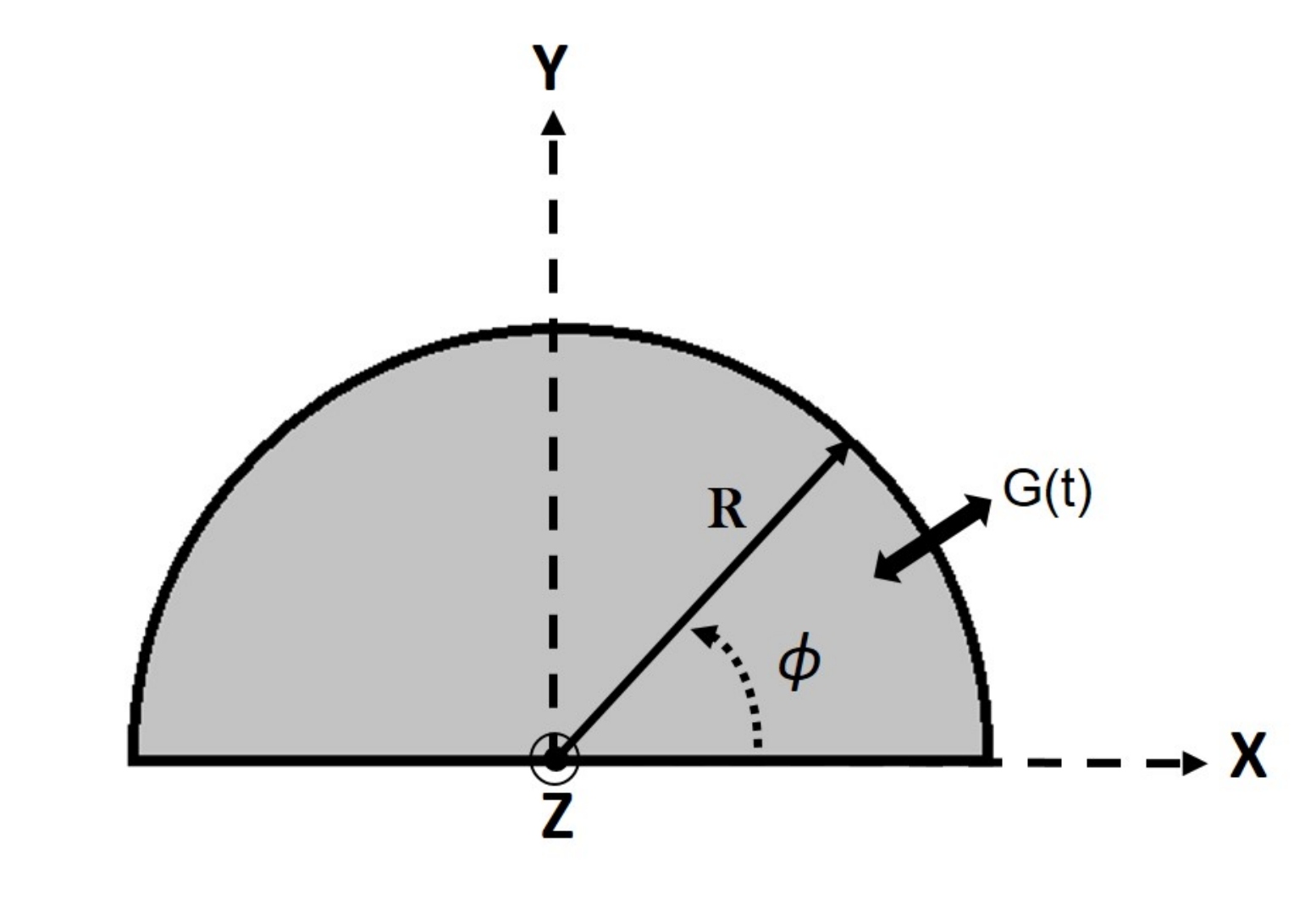}
\caption{Schematic view of the transverse cross-section of the water half-cylinder with the coordinate system used for the theoretical model.
} \label{schematic_theory}
\end{center}
\end{figure}

Excitation of standing waves on a water half-cylinder at the primary instability is sensitive to the radius of the cylinder. If the radius is too small or too large, it is difficult to make the vertical cross-section 
semicircular. Nearly, semi-circular cross-sections were in the range of 
$2.0~\rm{mm} \le R \le 4.5~\rm{mm}$. If the radius was increased more than $5.0$ mm, the boundary cross-section of the stationary water horizontal column was distorted and was no more semicircular. Developing an accurate theoretical model beyond $R > 5.0$ mm might be a formidable task. For water, density $\rho =1$ gm/cc and surface tension $\sigma =72$ dynes/cm. The acceleration due to gravity $g$ is taken as $980$ cm/s$^2$. The height (radius) of water half-cylinder was less than or comparable  to the capillary length of water $l_c \approx \sqrt{2 \sigma /(\rho g)} \sim 4$ mm, beyond which the gravity effects are expected to be prominent. Therefore, we chose the above mentioned range of radius. If the diameter (width of the column) is increased beyond $10.0$ mm, water bed spilled over the plate when the external driving was switched on. The height (vertical radius) was measured from the transverse direction using a vertical calibrated scale adjacent to the horizontal water column. Unlike a one-dimensional stretched string where the wavelength is decided for any chosen length, in this case longitudinal and azimuthal wave numbers may not fit simultaneously for arbitrary chosen $R$ and $L$. In the range of $2.0~\rm{mm} \le R \le 4.0~\rm{mm}$ and $ 10.0~\rm{cm} \le L \le 15.0~\rm{cm}$, we observed nice fitting of standing waves. For $R > 5.0~\rm{mm}$, the vertical cross-section was no more semicircular and the cylindrical modes were not observed. 

To identify the excited modes on the water half-cylinder, we make a simple model of various modes of the surface deformation. A cylindrical coordinate system is chosen with the $z$-axis coincident with the axis of water half-cylinder with origin at any point on the axis (see Fig.~\ref{schematic_theory}). The angle $\phi$ is measured from the line of intersection of the $r \phi$-plane (the vertical plane normal to the cylindrical axis) with the horizontal plate surface in counter-clockwise direction. Due to excitation of standing waves, the free surface of the water half-cylinder is now located at $r_\mathrm{s} (\phi, z ,t)$ $=$ $R + \zeta (\phi, z ,t)$, where $\zeta (\phi, z ,t)$ is the surface deformation.  In the experiment, the no-slip condition at the base and the hydrophobic paint outside the base of  water half-cylinder lead to  $\zeta (0,z,t)=\zeta (\pi, z, t)=0$. We therefore express the $m^{th}$ eigen mode of the surface deformation $\tilde{\zeta}_m (\phi, z, t)$ as
\begin{equation}
\tilde{\zeta}_m (\phi, z, t) = \zeta_{0m} \sin{(\omega t/2)} \sin{(m\phi)} \sin{(k_{m}z)},
\label{surface_modes}
\end{equation}
which is consistent with the boundary conditions of the experiment.
Here $\zeta_{0m}$ and $k_m = 2\pi/\lambda_m$ are the amplitude and experimentally observed wave number along the cylinder axis of an eigen mode with  azimuthal wave number $m$. 

\begin{figure}[]
\begin{center}
\includegraphics[height=!, width=8.5 cm]{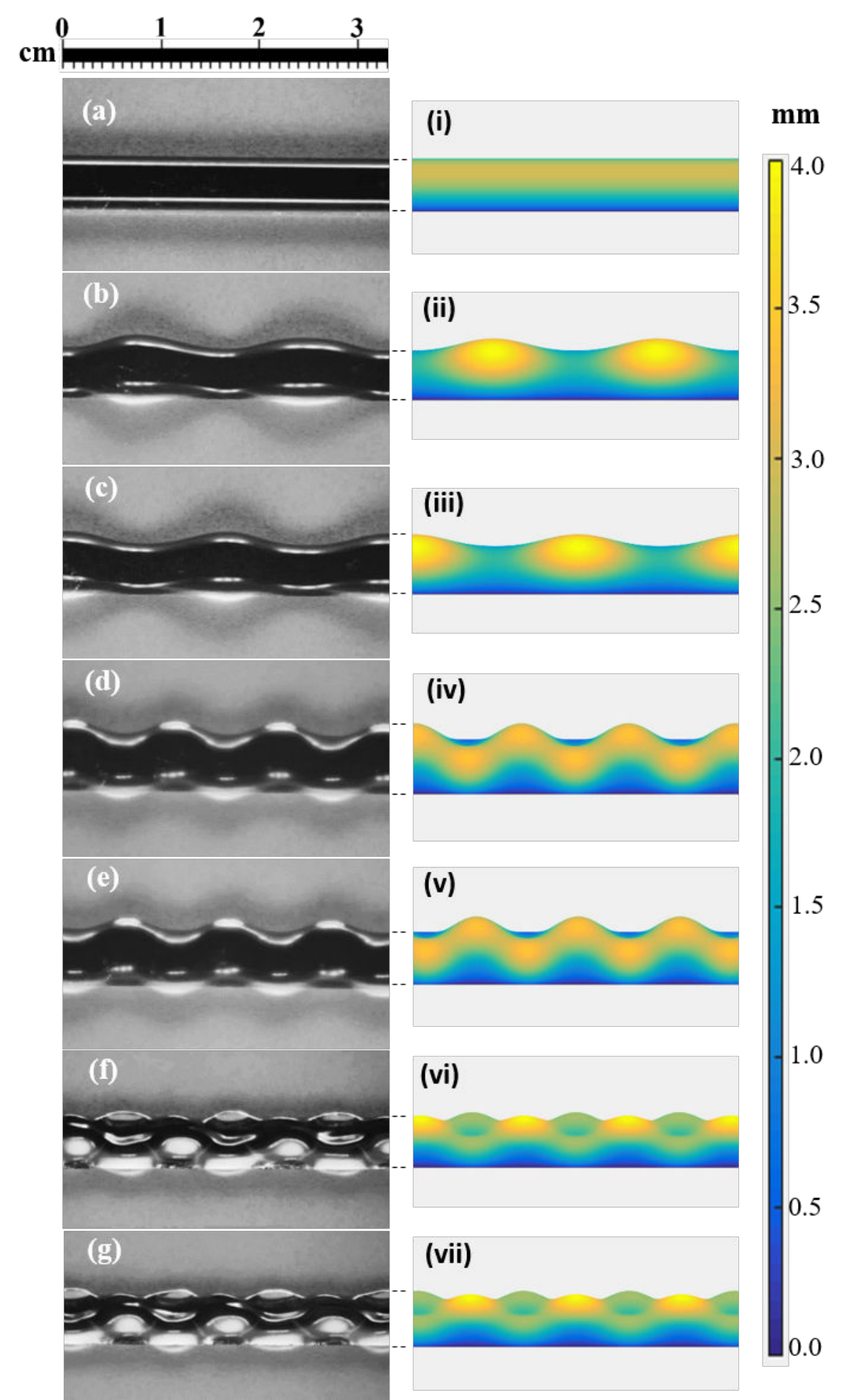}
\caption{(Color online) View from an angle of the experimentally observed stationary waves on a portion of water half-cylinder ($L=10.0\pm 0.1$ cm, $R=3.1\pm 0.1$ mm) (see the left column) and computer generated patterns (right column).  Images [(a) - (e)] are captured at angle $60^{\circ}$ and those in [(f) - (g)] are captured at angle $40^{\circ}$ from the horizontal. All other experimental conditions for these figures are the same as those in Fig. ~\ref{top_view}. For computer generated patterns camera angle are set same as those for the  corresponding experimental images. White portions in the experimentally captured images are due to reflections of light from the curved surface.
} \label{angular_snapshots}
\end{center}
\end{figure}

The images in the right column of Fig.~\ref{top_view} [(i)-(vii)] are the simulated surface deformation for different eigen modes.  The viewing angles are the same as those used in capturing the real images (see the left column).  However, there are no effects of light in the simulated images.  We have normalized the amplitude of the surface deformation, $\zeta_{0m}$, of the simulated images to the corresponding real ones.  Small dashed lines between the experimental and computer generated images mark the actual boundaries of the water half-cylinder on the plate. Figure~\ref{top_view}(i) shows water surface in the absence of the external driving. The yellow (light gray) and blue (dark gray) correspond to higher and lower elevations, respectively.

Figure~\ref{top_view}(ii)-(iii) displays the snapshots of the simulated surface deformation for eigen mode with $m=1$ at two instants separated by half of the wave period for driving frequency $f=20$ Hz. The  bulges due to surface deformation match precisely. Lower portions differ slightly from the corresponding real images of the experiment [Fig.~\ref{top_view}(b)-(c)] due to lighting effects. Figure~\ref{top_view}(iv)-(v)] shows the snapshots of the simulated surface at $f=60$ Hz due to the excitation of waves for $m=2$. The concave and convex bulges match well with corresponding images of the experiment [Fig.~\ref{top_view}(d)-(e)]. Ignoring the reflections of light from the curved free-surface, the simulated images look identical to the ones observed in the experiment.  Figure~\ref{top_view}(vi)-(vii) displays the images for $m=3$. The  yellow (light gray) regions and the bluish green (gray) regions surrounded by light yellow (light gray) correspond to mountains and valleys, respectively. They are quite similar to the fluid patterns observed at driving frequency $f = 90$ Hz [Fig.~\ref{top_view} (f)-(g)].  

Figure~\ref{top_view} dispalys the symmetry of the figure clearly and it also identifies various excited modes of the water half-cylinder. However, it does not clearly reveal the structures of the fluid half-cylinder under vibration. The fluid structures are better observed at viewing the patterns not from the top but at an angle.  The left column of Fig.~\ref{angular_snapshots} shows all the experimental images captured at different viewing angles with the horizontal direction. The images in the right column of Fig.~\ref{angular_snapshots} are computer generated images corresponding the images in the left column.  The viewing angle was set equal to $60^{\circ}$ for Fig.~\ref{angular_snapshots}(a)-(e).  The viewing angle was set  equal to $40^{\circ}$ for Fig.~\ref{angular_snapshots}(f)-(g). All other parameters to capture these images were exactly the same as those corresponding to the images given in the left column of Fig.~\ref{top_view}. Videos enclosed show different fluid patterns on the full length of a vibrating water half-cylinder. 

\begin{figure*}[]
\begin{center}
\includegraphics[height=!, width=17 cm]{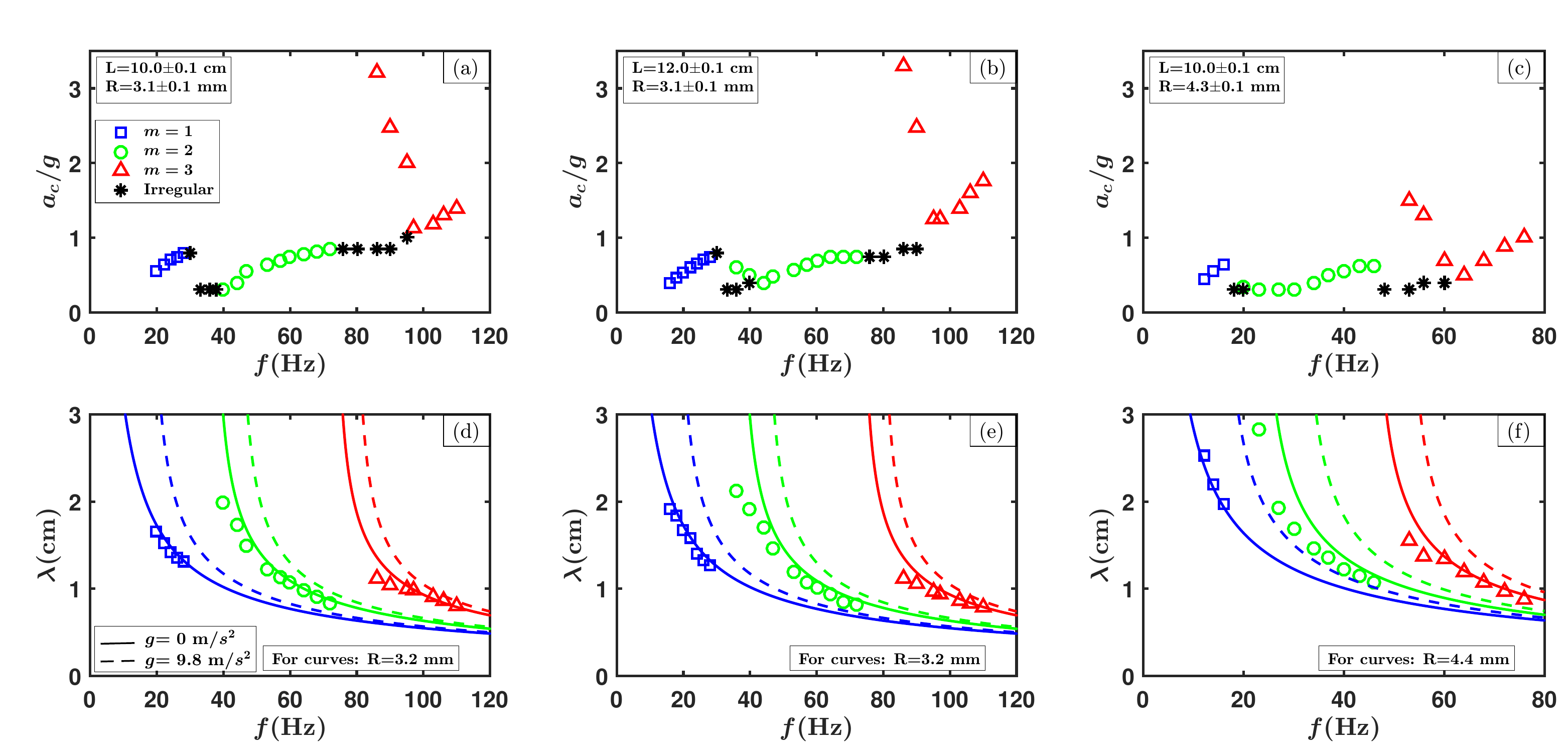}
\caption{(Color online) Thresholds [(a)-(c)] and dispersion curves [(d)-(f)] for sub harmonically excited stationary waves on water half-cylinders. Blue (black), green (light gray) and red (gray) symbols (curves) are experimental (theoretical) data points for waves with $m = 1$, $2$ and $3$, respectively. Black stars are experimental data points showing irregular spatio-temporal patterns.  
}
\label{dispersion_threshold}
\end{center}
\end{figure*}

Figure~\ref{dispersion_threshold}(a)-(c) shows the dimensionless threshold, 
$a_c/g$, as a function of driving frequency $f$ for the instability of  horizontal water half-cylinders with different $L$ and $R$ combinations. Blue (black) squares, green (light gray) circles and red (gray) triangles are experimental data points for sub harmonically excited stationary waves in the form of mirror symmetric half-beads ($m = 1$), waving half-cylinder with glide symmetry ($m = 2$) and complex knitting patterns with mirror symmetry ($m = 3$), respectively.  Black stars between $m=1$ and $m=2$ and also between $m=2$ and $m=3$ patterns [Fig.~\ref{dispersion_threshold}(a)-(c)] are experimentally observed data points when the water half-cylinder is in a state of frustration at primary instability. The location and the frequency windows of such states are sensitive to the radius, $R$, of the half-cylinder.  These frustrated spatio-temporal states may bifurcate to an ordered sate (stationary waves) at secondary instability at higher values of $a$ [see the upper set of red (gray) triangles]. These are large amplitude nonlinear states. The fluid patterns generally become irregular at higher values of $a$. Figure~\ref{dispersion_threshold}(d)-(f) displays the corresponding dispersion relations for sub harmonically generated stationary waves for the cases shown in Fig.\ref{dispersion_threshold}(a)-(c), respectively. Similar symbols correspond to the similar fluid patterns and size of the symbols includes the error bars in both directions.

\section{A theoretical model}
We now present a linear theory to understand the dispersion relation for excitation of capillary waves on a long horizontal water half-cylinder of radius $R$. As water viscosity is small, it is ignored. The determination of acceleration threshold requires a theory of viscous liquid, but the dispersion relation for low viscosity fluids may be computed using this theory. A cylindrical coordinate system is chosen as described earlier (in the paragraph preceding Eq.~\ref{surface_modes}). The pressure $P_0$ inside the  undeformed vibrating water half-cylinder is given by ${\boldsymbol{\nabla}} P_0 (r, \phi, t)$ $=$ $\rho  G(t)\hat{\boldsymbol{\mathrm{y}}}$ $=$ $-\rho(g + a \cos{\omega t})$ $\hat{\boldsymbol{\mathrm{y}}}$, where $\omega = 2\pi f$ and $\hat{\boldsymbol{\mathrm{y}}}$ is a unit vector along the vertical direction. The pressure jump across the static cylindrical fluid surface is equal to $\sigma/R$.  As the flat plate starts  oscillating, the cylindrical free surface is deformed, a liquid  particle on the curved surface experiences acceleration along its normal and tangential directions. However, the tangential component is compensated by the surface forces due to surface tension.  Since the bottom of water half-cylinder sticks to the horizontal plate due to no-slip condition, there is additional stress at the bottom when waves are excited. In addition, there is reaction of the plate on water half-cylinder, which is absent in a vertical liquid cylinder (jet) and a freely falling small water droplet. The combined effects of all these may lead to an effective acceleration along the radial direction. The experimentally observed standing waves appear as periodic expansion and contraction of the water half-cylinder. So we assume the effective acceleration due to vibrating plate in radial direction.
Recently, the dispersion relation for a vibrating spherical liquid drop~\cite{Adou_Tuckerman_2016} was determined by considering  the external acceleration $\boldsymbol{\mathrm{G}}$ only along the radial direction. We therefore make a simplification by assuming $\boldsymbol{\mathrm{G}}(t) = G(t)\hat{\boldsymbol{\mathrm{r}}}$, in the perturbation equations. As soon as waves are excited, the velocity field $\boldsymbol{\mathrm{u}} (r, \phi, z, t)$ develops in the water half-cylinder. It may be written as 
$\boldsymbol{\mathrm{u}}$ $=$ ${\boldsymbol{\nabla}}\Phi$, where $\Phi (r, \phi, z, t)$ is the velocity potential. The modified pressure in water half-cylinder is $P (r, \phi, z, t) = P_0 (r, \phi, t)+ p(r, \phi, z, t)$, where $p$ is the deviation in pressure field from $P_0$ due to instability.  The incompressibility condition of the liquid leads to Laplace equation for $\Phi$, i.e., $\nabla^2\Phi = 0$. The kinematic condition at the curved surface reads as:
\begin{equation}\label{kinematic}
\frac{\partial{\zeta}}{\partial{t}}=u_r|_{r=R}=\frac{\partial{\Phi}}{\partial{r}}|_{r=R}.
\end{equation}
The pressure jump across the free surface now reads as:
\begin{equation}\label{eq:normal_stress}
\frac{\partial{\Phi}}{\partial{t}}|_{r=R} = G(t)\zeta + \frac{\sigma}{\rho}\left( \frac{\zeta}{R^2}
+\frac{1}{R^2}\frac{\partial^2{\zeta}}{\partial{\phi^2}}
+\frac{\partial^2{\zeta}}{\partial{z^2}}\right).
\end{equation}
The velocity potential $\Phi$  and the surface deformation $\zeta$ are expanded as:  
\begin{equation}\label{expansions}
(\Phi; \zeta)= \sum_{m=1}^{\infty}\left(\Phi_m (t) I_m(kr);\zeta_m (t)\right) \sin{(m\phi)} e^{ikz},
\end{equation}
where $I_m(kr)$ is the $m^{th}$ order modified Bessel function of the first kind. The integer $m$ is the same azimuthal wave number used in Eq.~\ref{surface_modes}. In a liquid half-cylinder on a flat surface, where the lower surface is always in contact with non-painted surface, only sinusoidal azimuthal modes are possible. This considers the deformation of the curved surface of the half-cylinder whose flat bottom of thickness $2R$ remains intact. This is actually the situation in experiments due to weak viscosity of water. Insertion of Eq.~\ref{expansions} in Eq.~\ref{eq:normal_stress} and use of Eq.~\ref{kinematic} yield a Mathieu equation for $\zeta_m$: 
\begin{equation}\label{eq:mathieu}
\ddot{\zeta}_m+\omega_{m}^2 \left[ 1 + ( a/a_{m}) \cos{(\omega t)}\right]\zeta_m =0,
\end{equation}
where $\omega_{m}^2$ $=$ $\frac{k I'_m(kR)}{I_m(kR)}\left[ g + \frac{\sigma}{\rho R^2 }(k^2R^2+m^2-1)\right]$,  $a_{m}$ $=$ $\frac{I_m(kR)}{kI'_m(kR)}\omega_m^2$.  The expression for $\omega_m$ is exactly the same as the dispersion relation for capillary instability of a vertical liquid cylinder~\cite{Rayleigh} in the absence of gravity.  Floquet expansion of $\zeta_m (t)$ is given as:
\begin{equation}\label{Floquet:expansion}
\zeta_m (t) = e^{(s+i\alpha \omega ) t} \sum_{n=-\infty}^{\infty} \zeta_m^{(n)} e^{i n\omega t},
\end{equation}
where $s$ is the growth rate and $\alpha$ is the Floquet exponent. Insertion of Eq.~\ref{Floquet:expansion} in the Mathieu equation (Eq.~\ref{eq:mathieu}) then leads to a difference equation:
\begin{equation}
 A_m^{(n)}\zeta_m^{(n)} = a \left( \zeta_m^{(n-1)}+\zeta_m^{(n+1)}\right),
\end{equation}
where 
\begin{equation}
A_m^{(n)} = -\frac{2 I_m(kR)}{kI'_m(kR)}\left[ \omega_m^2-(n+\alpha)^2 \omega^2\right].
\end{equation}
This difference equation can be converted to an eigenvalue matrix equation~\cite{Kumar_1996}. Real and positive eigenvalues of the matrix can be determined as a function of wave number $k$ for fixed values of $\omega$ and $m$. This gives marginal ($a-k$) curve (the growth rate, $s =0$ ) for given values of the azimuthal wave number $m$. The stationary waves are excited sub harmonically only when $\omega_{m} = \omega/2$.  We set $\alpha =1/2$ and $s=0$, as we are interested in sub harmonically excited waves at the instability onset.  The minimum of the $a-k$ curve gives the instability threshold and the critical wave number for sub harmonically excited stationary waves for fixed values of $\omega$ and $m$. The dispersion curves can be computed by varying $\omega$ in small steps and finding the critical values of $k$ for a fixed value of $m$. Solid (dashed) curves are dispersion relations computed for water half-cylinders of different radii without (with) $g=980$ cm/s$^2$ for different values of $m$ [see, Fig.~\ref{dispersion_threshold}(d)-(f)]. Blue (black), green (light gray) and red (gray) colored curves correspond to $m=1$, $2$ and $3$, respectively. The primary instability wavelengths match nicely with theoretical predictions even at higher frequencies with the continuous curves computed from the theory without $g$. Moreover, slight deviations are observed for wavelengths $\lambda$ (for $m = 1$ and $2$) greater than $1.5$ cm. The critical wavelength for water waves is around $2\pi \sqrt{\sigma/(\rho g)} \sim 1.7$ cm, beyond which the gravity effects would be considerable. The inclusion  of gravity in the model makes the fit worse. The instability observed is therefore seems to be primarily curvature influenced. The slight deviations of experimental data points, from the theoretical dispersion curves at higher driving amplitudes (for $m = 3$) are for stationary waves observed at the secondary instability and they naturally do not match so well due to nonlinear effects. These deviations could be seen even for wavelengths less than $1.5$ cm. The understanding of which may require a nonlinear theory perhaps with viscosity.    

Capillary waves with frequency three and half times  the driving frequency (superharmonics) were observed in micro air-water meniscus~\cite{Xu-Attinger_2007} in water under pressure oscillations. We did not observe superharmonically excited waves. In our case all observed patterns were  subharmonically generated. Here, the azimuthal wave number $m$ took different values (e.g., $m$ $=$ $1, 2, 3$) corresponding to structures with different symmetries on water half-cylinder. For each value of the azimuthal wave number $m$, we have one  Mathieu equation~(Eq.~\ref{eq:mathieu}). The azimuthal number does not play any role in the Mathieu equation used for the superharmonic waves in air-water meniscus~\cite{Xu-Attinger_2007}.  

\section{Conclusions}
A horizontal water half-cylinder under vertical vibration becomes unstable and leads to only non-axisymmetric ($m\neq0$) subharmonic stationary waves, which are new and qualitatively different from the axisymmetric patterns ($m=0$) of Savart-Plateau-Rayleigh instability in vertical liquid columns and jets. These curvature influenced waves possess either mirror symmetry ($m$ odd) or glide symmetry ($m$ even), which distinguish them from the waves observed in a planer Faraday system  or in spherical liquid drops under vertical driving. Excitation of different azimuthal modes at different values of the driving frequency leads to different fluid patterns. Water half-cylinder oscillates as a periodic chain of half-beads with mirror reflection ($m=1$) at lower frequencies. It becomes a wavy half-cylinder with glide symmetry ($m=2$) at slightly higher frequencies. It shows a complex knitting pattern with mirror symmetry ($m=3$)water at relatively higher frequencies. The water half-cylinder is in a state of frustration, if standing waves do not fit on the half-cylinder of a finite length. The results presented here have possible potential applications in  material processing, microfluidic flows, fluid atomization, coating and drug mixing to name a few.

\begin{acknowledgements}
{\noindent Partial support from SERB, India \\through Project Grant No.
EMR/2016/000185 is acknowledged. Authors acknowledge fruitful suggestions from anonymous referees, which improved the manuscript.} 
\end{acknowledgements}

\end{document}